# Cosmology and Science Education: Problems and Promises[*]

Helge Kragh[**]

**Abstract** Cosmology differs in some respects significantly from other sciences, primarily because of its intimate association with issues of a conceptual and philosophical nature. Because cosmology in the broader sense relates to the students' world views, it provides a means for bridging the gap between the teaching of science and the teaching of humanistic subjects. Students should of course learn to distinguish between what is right and wrong about the science of the universe. No less importantly, they should learn to recognize the limits of science and that there are questions about nature that may forever remain unanswered. Cosmology, more than any other science, is well suited to illuminate issues of this kind.

## 1. Introduction

Whether majoring in science or not, students at high school and undergraduate university level are confronted with issues of cosmology, a subject which has only attracted a limited amount of attention in the context of science education (Kragh 2011a). It is important that when students are introduced to cosmology, this is done correctly not only in the technical sense but also in a conceptual sense. As shown by several studies, misconceptions abound in both areas. They include some of the philosophical aspects that are so closely intertwined with cosmology in the wider sense and to which a large part of cosmology's popular appeal can be attributed. These aspects need to be addressed and coordinated with the more standard, scientific aspects. In this respect it is often an advantage to refer not only

---

to the modern big bang theory but also to older developments that may illuminate modern problems in cosmology in a simple and instructive manner.

Following a brief discussion of the development of cosmology as a science, the article focuses on various conceptual misunderstandings that are commonly found in students' ideas about modern cosmology. Some of these misconceptions are of a philosophical nature, for example related to the concept of the universe and its supposed birth in a big bang. By taking issues of this kind seriously, students will hopefully be brought to reflect on the limits of science and adopt a critical attitude to what scientific cosmology can tell us about the universe.

## 2. Early Cosmology: Lessons for Science Education

According to the view of most physicists and astronomers, and also according to some historians of science (Brush 1992), cosmology became a science only in the twentieth century. Some will say that the supposed turn from 'philosophical' to truly scientific cosmology only occurred with the discovery of the cosmic microwave background radiation in 1965, while others date the turn to Edwin Hubble's insight in the late 1920s of the cosmological significance of the galactic redshifts. Others again suggest that the turning point is to be found in Einstein's cosmological model of 1917 based on his general theory of relativity.

The widely held opinion that there was no scientific cosmology – scientific in more or less the modern sense of the term – before Einstein and Hubble entered the stage is reflected in most introductory textbooks in physics and astronomy. The general structure of these books is to start with the solar system and then proceed to stars and galaxies, ending with the universe as a whole. The chapters on cosmology are usually restricted to post-1920 developments (Krauskopf & Beiser 2000). Although earlier developments are sometimes included, then it occurs in sections that appear separate from the account of modern cosmology and are typically placed in the beginning of the book. For example, the epic confrontation between the Aristotelian-Ptolemaic universe and the heliocentric world system during the so-called Copernican revolution is a classic theme in the teaching of physics and astronomy, where it is often presented as a methodological case study. On the other hand, textbooks and similar teaching materials rarely refer to other parts of the rich history of cosmological thought, for which teachers and students must look up the literature written by historians of science (North 1994, Kragh 2007). The exception to this state of affairs is Olbers' famous paradox of the dark night sky, dating from 1826 but with roots back to Kepler, which can be found in most textbooks.

Although modern cosmology dates in most respects from the early part of the twentieth century, it does not follow that earlier theories about the universe were not scientific. The cosmos of the ancient Greeks was very different from ours, yet Ptolemy's cosmology was basically scientific in so far that it was a mathematical model that rested on observations and had testable consequences. At any rate, there are good reasons to include aspects of pre-Einsteinian cosmology also in the context of science education. For one thing, students should be aware of this earlier development for general cultural reasons. Moreover, the earlier history of cosmology provides many more examples of educational relevance than just the one of the Copernican revolution. Although Michael Crowe's two books on theories of the universe are not ordinary textbooks, they are based on his very extensive experience with teaching history of astronomy and cosmology at the University of Notre Dame (Crowe 1990, Crowe 1994). They are of value to the teacher of introductory astronomy courses because they include a large amount of primary sources from Ptolemy to Hubble that can be easily used in the classroom. Moreover, Crowe (1994) includes laboratory exercises related to the studies of the nebulae by William Herschel in the late eighteenth century and by the Earl of Rosse in the mid nineteenth century.

To illustrate the relevance of earlier cosmological thought in science education, consider the discussion in the thirteenth century concerning the possibility of an eternal yet created universe. The discussion was abstract and philosophical, not scientific, but it is nonetheless of relevance to problems of modern cosmology because it led the scholastic thinkers to scrutinize the concept of creation in a sophisticated way that went beyond the identification of creation with temporal beginning (Carroll 1998, see also Sect. 6.4). As another example one might point to the difficult problem of spatial and material infinity as it turned up in Newton's correspondence with Richard Bentley in the early 1690s. Both Bentley and Edmund Halley mistakenly believed that in an infinite stellar universe each star would be attracted by equal forces in any direction and therefore be in a state of equilibrium. The belief is intuitively convincing and probably shared by most students, but Newton knew better. As he pointed out, two infinities do not cancel. The case is well suited to discuss with students the tricky problems of infinities that appear no less prominently in modern cosmology than they did in the past.

Students should also be aware that the fundamental distinction between realism and instrumentalism, an important issue in the discussion of the nature of science (Campbell 1998), does not only turn up in microphysics but also in cosmology. After all, the universe is no less unobservable than are quarks and superstrings. No one has ever observed the universe and no one will ever do so, so how can we know that the universe exists? The realist will claim that 'the universe' designates an entity that exists independently of all cosmological enquiry, while the instrumentalist considers it a concept that can be ascribed a meaning only in a pragmatic sense, as it is a construct of cosmological theory. The tension between the two opposite views can be followed through much of the history of cosmology, from Ptolemy's world system to the modern multiverse, and from a teaching point of view it may sometimes be an advantage to refer to older sources rather than to modern examples. To illustrate cosmological or astronomical antirealism with regard to theories, one may read passages of Stephen Hawking (a positivist and instrumentalist), but the same point is brought home, and with greater clarity, by Andreas Osiander's notorious preface to Copernicus' *De Revolutionibus*.

**3. Patterns in the Development of Modern Cosmology**

To the extent that practicing scientists are familiar with philosophical theories of science, the theories are often limited to the views of Karl Popper and Thomas Kuhn. The ideas of these two philosophers are also likely to be the only ones that students will meet, either explicitly or implicitly, in physics and astronomy courses.

While historians agree that Kuhn's theory of scientific revolutions does not in general fit very well with the actual history of science, the history of cosmology yields some support for the notion of paradigm-governed science and revolutionary changes, if not in the radical sense originally proposed by Kuhn (Kragh 2007, pp. 243-245). In both the older and the modern history, there are several cases of beliefs and traditions that formed the nearly unquestioned framework of cosmological thinking and hence had the character of paradigm. Thus, until about 1910 it was generally believed that the stellar universe was limited to the Milky Way. As the astronomy writer Agnes Clerke asserted, 'No competent thinker, with the whole of the available evidence before him, can now, it is safe to say, maintain any single nebula to be a star system of coordinate rank with the Milky Way' (Clerke 1890, p. 368). She added: 'With the infinite possibilities beyond, science has no concern'.

Likewise, until 1930 the static nature of the universe as a whole was taken for granted. Current cosmology is solidly founded on Einstein's theory of general relativity theory and some kind of big bang scenario, elements that are largely beyond discussion and conceived as defining features of cosmological theory. Yet, although it may be tempting to characterize these beliefs as paradigmatic, they are so in a different sense from what Kuhn spoke of in his classical work of 1962, *The Structure of Scientific Revolutions*. First of all, there is no indication of radical incommensurability gaps in the development that led from the static Milky Way universe to the current standard model of big bang cosmology.

The applicability of the Kuhnian model to the case of modern cosmology has been investigated by Marx and Bornmann (2010) by means of bibliometric methods. They examine what they misleadingly call 'the transition from the static view of the universe to the big bang in cosmology', a process that supposedly occurred in the mid-1960s when the steady state model was abandoned in favour of the hot big bang model. (In reality, the transition from a static to a dynamic universe occurred in the early 1930s and was unrelated to ideas about a big bang.) As indicated by bibliometric data the emergence of the victorious big bang model in the 1960s marked a drastic change in cosmology, if not a sudden revolution.[1] Based on citation analysis the two authors suggest that if there were a paradigm shift, it was a slow process ranging from about 1917 to 1965 – which cannot reasonably be called a paradigm shift in Kuhn's sense.

---

[1] The number of publications on cosmology grew dramatically in the 1960s, apparently an indication of the revolutionary effect caused by the standard big bang theory (Kaiser 2006, p. 447, Marx & Bornmann 2010, p. 543). However, the growth is in some respect illusory, as the number of publications in the physical and astronomical sciences as a whole grew even more rapidly. While cosmology in 1950 made up 0.4% of the physics research papers, in 1970 the percentage had shrunk to a little less than 0.3% (Ryan & Shepley 1976). Numerical data can be presented in many ways, sometimes resulting in opposite messages.

In an earlier paper Shipman (2000) found that nearly half of his sample of astronomers had never heard of Kuhn and that an additional third was only vaguely familiar with him. Of those who were aware of Kuhn's philosophy, several responded that it informed their teaching and consequently was of value in the classroom. One respondent said: 'I think changing paradigms are so obvious in astronomical history that it goes almost without saying that his work is interesting to an astronomer, but I never thought to actually make a big deal of it in class' (Shipman 2000, p. 165). Whereas some astronomers found Kuhn's model to be helpful in understanding the development of the astronomical sciences, none of them thought it was relevant to their research or had an impact on modern astronomy and cosmology. As one astronomer responded, 'Kuhn … has no effect on the way science is done' (p. 169).

In this respect, the case of Popper is rather different as his falsificationist philosophy of science has exerted a strong and documented influence on the astronomical and cosmological sciences and continue to do so (Sovacool 2005, Kragh 2013). Although most cosmologists are only superficially acquainted with Popper's ideas, which they tend to use in a simplified folklore version, they often invoke them as a guide for constructing and evaluating theories. This is evident from the modern controversy over the multiverse, and it was just as evident in the past, when Popperian standards played an important role in the debate between the steady state theory and the class of relativistic evolution theories (Kragh 1996, pp. 244-246). Hawking has little respect for philosophy, but in his best-selling *A Brief History of Time* he nonetheless pays allegiance to the views of Popper:

> Any physical theory is always provisional, in the sense that it is only a hypothesis: you can never prove it. ... On the other hand, you can disprove a theory by finding even a single observation that disagrees with the predictions of the theory. As philosopher of science Karl Popper has emphasized, a good theory is characterized by the fact that it makes a number of predictions that could in principle be disproved or falsified by observation. (Hawking 1989, p. 11).

Influential as Popperianism is in cosmological circles, the influence is mostly limited to the popular literature and general discussions of a methodological nature. As it is the case with Kuhn, Popper's name very rarely appears in research papers. Perhaps more surprisingly, the same seems to hold for elementary textbooks in astronomy and cosmology. On the other hand, the influence of a philosopher may be visible even though his or her name is missing. Thus, in a brief methodological section astronomy author Karl Kuhn writes: 'A theory of science must be able to be shown to be wrong. A theory must be testable. Every theory must be regarded as tentative, as being only the best theory we have at present. It must contain within itself its own possibility of destruction' (Kuhn 1998, p. 557). It is then up to the teacher whether Popper should be named or not.

**4. Conceptions and Misconceptions of Cosmology**

Most of the misconceptions about cosmology commonly found among students concern the two fundamental concepts of the expanding universe and the big bang. The two concepts are closely connected, but the precise connection between them is often misconceived.

*4.1. The Expanding Universe*

The standard tradition in introductory astronomy and physics textbooks dealing with cosmology is understandably characterized by an emphasis on observations rather than theory. Observations are used as arguments for new concepts and often presented in a historical context. Expositions typically start with two important and connected observations from the early decades of the twentieth century, Melvin Slipher's discovery in the 1910s of galactic redshifts and Hubble's conclusion from 1929 of a linear relationship between the redshifts and the distances of the galaxies. Both of these historical cases are easily comprehended and can, moreover, be turned into students' exercises by providing the students with the data used by the two astronomers, or by using the students' own data found with a 'simulated telescope' (Marschall, Snyder & Cooper 2000). From the Hubble relation there is but a small step to the expanding universe. Almost without exception textbooks and popular expositions illustrate the expansion of space by means of the inflating-balloon analogy, which may also be used to introduce the notion of curved space such as applied in relativistic cosmology. This standard analogy – to 'imagine the nebulæ to be embedded in the surface of a rubber balloon which is being inflated' – was first suggested by Arthur Eddington in 1931, shortly after the expansion of the universe had been recognized (Eddington 1931).

Although there may be but a small step from the Hubble relation to the expanding universe, the step is real and should not be ignored. Students may be told that the expansion of the universe is an observational fact, but this is not quite the case. We do not *observe* the expansion, which does not follow from the data of either Hubble or later observers. As Hubble was keenly aware of, it takes theoretical assumptions (such that the redshifts are due to a Doppler effect) to translate the measured redshifts into an expansion of the universe. It is quite possible to accept the redshift-distance relation and, at the same time, maintaining that the universe is static, such as many scientists did in the 1930s and a few still do. In fact, Hubble, a cautious empiricist, never concluded that the universe is in a state of expansion. What is 'commonly known' and stated in many textbooks and articles, namely that 'The expansion of the universe was discovered by Edwin Hubble in 1929' (Lightman & Miller 1989, p. 135), is just wrong. Hubble did not discover the expansion of the universe and he never claimed that he did (Kragh & Smith 2003).

There is a tendency in textbooks, perhaps understandable from a pedagogical perspective, to simplify and dramatize discoveries. For example, one textbook presents Hubble's discovery of the redshift-distance relation as follows: 'The law was published in a 1929 paper on the expansion of the universe. It sent shock waves through the astronomical community' (Kuhn 1998, p. 512). However, it is only in retrospect that Hubble's paper was about the expansion of the universe, and it did not initially create a stir in either the astronomical or the physical community. According to the *Web of Science*, in the years 1929-1930 it received only three citations in scientific journals.

A much better candidate for the discoverer of the expanding cosmos is the Belgian pioneer cosmologist Georges Lemaître, who in a work of 1927 clearly argued that the universe was expanding and even calculated the quantity that came to be known as the Hubble constant (Holder & Mitton 2012). Contrary to Hubble, Lemaître was fully aware that the measured galactic redshifts are not due to a Doppler effect of galaxies flying through space, but must be interpreted as the stretching of standing waves due to the expansion of space, that is, as a relativistic effect. As he explained, if light was emitted when the radius of curvature of the closed universe was $R_1$ and received when it had increased to $R_2$, the 'apparent Doppler effect' would be given by $\Delta\lambda/\lambda = R_2/R_1 - 1$. The important difference between the Doppler explanation and the relativistic expanding-space explanation can be illustrated in a simple way by means of the balloon analogy (Lotze 1995). One should distinguish between the expansion of space and the expansion of the material universe, such as most textbooks do. It is much easier to comprehend galaxies moving apart, as were they flying through space, but it is more correct to conceive space as expanding and the galaxies changing their relative positions because of the expansion of space. The counterintuitive notion of an expanding empty space, such as implied by the model first studied by Willem de Sitter in 1917, illustrates the difference between the two explanations.

As documented by many studies, the expansion of the universe is not well understood, if understood at all, by either the general public or general science students. Comins (2001) discusses a large number of astronomical and cosmological misconceptions, why they are held and how to correct them.[2] Unfortunately, when it comes to the history of cosmology he expresses several misconceptions of his own, including that Einstein, because he included the cosmological constant in his 1917 cosmological model, 'missed the opportunity to predict that the universe expands' (p. 162). This common misunderstanding is easily seen to be unfounded, for other reasons because the cosmological constant was part of Lemaître's expanding model of 1927 based on Einstein's equations. Moreover, Einstein did not introduce the cosmological constant to keep his universe from expanding, but to keep it from collapsing. In short, a cosmological model may describe an expanding universe whether or not it includes a non-zero cosmological constant.

Asked whether the universe is systematically changing in size or remaining about the same size, nearly 60% of 1111 interviewed American adults offered the last response. According to the survey conducted by Lightman and Miller (1989) only 24% of the respondents said that the universe is expanding. Later large-scale surveys of students following introductory astronomy courses confirm that they have difficulties with the expanding universe and other concepts of modern cosmology. Only a minority of the students revealed a reasonably correct understanding of the meaning of the 'expansion of the universe', and a sizeable minority denied that the universe is increasing in size. Instead they suggested that the phrase was a metaphor for how our knowledge of the universe has increased over time (Wallace, Prather & Duncan 2012). One student answered that the expanding universe is an expression for stars and planets moving away from a central area in the universe, if not necessarily from the Earth (Wallace, Prather & Duncan 2011).

---

[2] See also Comins' website on 'Heavenly errors' that includes nearly 1,700 common misconceptions that students and other people have about astronomy and cosmology. Among them are that the universe has stopped expanding, that there is a centre of the universe, and that all galaxies are moving away from the Earth (http://www.umephy.maine.edu/ncomins/).

Another question that often causes confusion is *what* takes part in the expansion. Although the expansion is 'universal', it does not refer to everything. Objects that are held together by other forces than gravity, such as electromagnetic and nuclear forces, remain at a fixed physical size as the universe swells around them. Likewise, objects in which the gravitational force is dominant also resist the expansion: planets, stars and galaxies are bound so strongly by gravitational forces that they are not expanding with the rest of the universe. There is no reason to fear that the distance of the Earth from the Sun will increase because of the cosmic expansion, although worries of this kind are not uncommon (Lightman & Miller 1989). In the survey conducted by Prather et al. (2003), 10% of the students thought that the expansion of the universe has terrestrial consequences, including the separation of the continental plates. Nor is our Local Group of galaxies expanding. The Andromeda Galaxy, for example, is actually approaching the Milky Way, causing a blueshift rather than a redshift. (In 1912, Slipher concluded that the Andromeda Galaxy approached the Sun, only subsequently to realize that it was an exception to the general pattern of galactic redshifts.) On the other hand, on a cosmological scale all matter is rushing apart from all other matter at a speed described by Hubble's law, $v = Hr$, where $H$ denotes the Hubble parameter or 'constant'. Since the Hubble time $1/H$ is an expression of the age of the universe, $H$ it is not really a constant but a slowly decreasing quantity.

There are other and more complex ways in which the expansion of the universe can be misconceived, some of them relating to the magical limit of the recession velocity apparently given by the speed of light $c$ (Davis & Lineweaver 2004, Ellis 2007, pp. 1214-1216). Students learn that nothing can move faster than the speed of light, which is a fundamental postulate of the theory of relativity. But according to Hubble's law the recession velocity keeps increasing with distance, implying that beyond the Hubble distance $c/H$ the velocity will exceed the speed of light. Can receding galaxies really cross this limit? If they do, will they then become invisible because their redshifts become infinite? In spite of the apparent contradiction with Einstein's postulate, superluminal recession velocities do not violate the theory of relativity. As Lemaître emphasized in 1927, the recession velocity is not caused by motion *through* space but by the expansion *of* space. According to general relativity theory, redshifts do not relate to velocities, as they do in the Doppler description (both classically and in special relativity), and the redshifts of galaxies on the Hubble sphere of radius $c/H$ will not be infinite.

Not only can the universe, or space, expand faster than the speed of light, we can also observe objects that recede from us with speeds greater than this limit. Students may believe that since the universe came into being 13.8 billion years ago, the most distant objects are 13.8 billion light-years away, but in that case they think in terms of a static universe. Since distances between faraway galaxies increase while light travels, the observability of galaxies is given by the look-back time, which is the time in the past at which light now being received from a distant object was emitted. As a result, the farthest object we can see is currently about 46 billion light-years away from us, receding with more than six times the speed of light, and that even though the universe is only 13.8 billion years old. The size of the observable universe is not given by the Hubble sphere but by the cosmic particle horizon beyond which we cannot receive light or other electromagnetic signals from the galaxies.

*4.2. The Big Bang*

Having digested the notion of expanding space, the next crucial concept that students need to be introduced to, the idea of the big bang, is often presented as a simple consequence of the cosmic expansion.[3] After all, if the distances between galaxies (or rather galactic clusters) increase monotonically, apparently there must have been a time in the past when all galaxies were lumped together. This inference is facilitated by the balloon analogy, where the airless balloon corresponds to the original universe before expansion. However, the inference is more seductive than correct. The argument from expansion to big bang may be pedagogically convincing, but it is not supported by either logic or the history of science. If there were such a necessary connection, how is it that while the majority of astronomers in the 1930s accepted the expansion of the universe, practically no one accepted the idea of an explosive origin?

In the version of the 'primeval atom' hypothesis the idea of a big bang was first suggested by Lemaître in 1931 – not in his 1927 paper, as is often stated – but it took many years until the hypothesis was taken seriously. The hypothesis was independently revived and much improved by George Gamow and his collaborators in the late 1940s, but even then it failed to win much recognition (Kragh 1996, pp. 135-141). Remarkably, from 1954 to 1963 only a single research paper was published on the big bang theory. During most of the period from about 1930 to 1960, the favoured theory of the evolution of the universe was the Lemaître-Eddington model according to which the universe had evolved asymptotically from a static Einstein state an infinity of time ago. This kind of model is ever-expanding but with no big bang and no definite age.

---

[3] The undignified name 'big bang' was coined by Fred Hoyle in a BBC radio programme of 1949, but neither Hoyle nor other scientists used it widely until the late 1960s. Contrary to what is often said (e.g., Marx & Bornmann 2010, p. 454), the phrase did not catch on either among supporters or opponents of the exploding universe. Hoyle belonged to the latter category, and it generally thought that he coined the name as a way of ridiculing the theory, but this is hardly the case. The first scientific paper with 'big bang' in its title appeared only in 1966.

Teachers presumably want their students to accept the big bang theory, but not to do it by faith or authority. To convince students that the big bang really happened they need to provide good reasons to believe in it, which primarily means observational and other empirical evidence. In this respect the students may be compared to the majority of astronomers and physicists who still in the 1950s resisted the idea of a big bang, basically because they lacked solid empirical evidence for the hypothesis. As the sceptics pointed out, quite reasonably, if our current universe has evolved from a very small and extremely dense and hot state several billions years ago, there must presumably still be some traces or fossils from it. If no such traces can be found, we have no reason to believe in the big bang and nor is there any possibility of testing the hypothesis.

An additional reason for the cool reception of the big bang theory was that according to most of the models, the calculated age of the universe came out embarrassingly small, much smaller than the age of the stars and smaller than even the age of the Earth. A universe that is younger than its constituent parts is of course ruled out for logical reasons. The age problem is mentioned in some astronomy textbooks, but not always historically correct. According to Arny (2004, p. 517), the age of Lemaître's primeval-atom universe was 2/3 times the inverse Hubble constant, which at the time, when Hubble's value $H$ = 500 km/s/Mpc was generally accepted, corresponded to only 1.2 billion years. The reference should be to the Einstein-de Sitter model of 1932, which assumed a flat space and a zero

16cosmological constant. Lemaître, on the other hand, assumed a positive cosmological constant by means of which he was able to avoid the age problem.

It is all-important that some kind of fossil is left over from the cosmic past, which otherwise would be inaccessible to us and therefore just a postulate one can believe in or not. It would have the same questionable ontological status as other universes in modern multiverse theories. In evidence-based courses in physics and astronomy students come to understand and accept the big bang picture by means of empirical evidence such as the cosmic microwave radiation and the abundance of helium in the universe. What matters is not so much the right scientific belief as it is to be able to justify these beliefs and distinguish them from ideas that are not adequately supported by evidence (Brickhouse et al. 2000, Brickhouse et al. 2002). Students learn that a theory must necessarily be supported by evidence and also that evidence depends on and is only meaningful in relation to the theory in question. The way students learn to accept the big bang corresponds to some extent to the historical situation in the period from about 1948 to 1965.

The celebrated discovery of the cosmic microwave background killed the already weakened rival steady state theory and turned the big bang theory into a successful standard theory of the universe.[4] Although the best known of the cosmic fossils, the microwave background is not the only one and nor was it the most important in the historical development of cosmology. It may be less well known that the distribution of matter in the universe provides us with another and more easily accessible fossil. None of the 219 students questioned by Bailey et al. (2012) referred to the chemical composition of the universe as evidence for the big bang, while 32 mentioned the expansion and three the cosmic microwave background as evidence.

---

[4] The classical steady state theory was abandoned half a century ago and for this reason is mainly of historical interest. On the other hand, from a methodological and also an educational point of view it is an instructive example of how an attractive theory with great predictive power was eventually shot down by new observations. In addition, it illustrates the aesthetic and emotional appeal of a cosmological theory, a phenomenon which is not restricted to the past. While Kuhn (1998, p. 555) covers the essence of the steady state theory, other textbook authors choose to ignore it (Krauskopf & Beiser 2000).

The hypothesis that the distribution of matter reflects the cosmic past was first proposed in the late 1930s, when the first reliable data of the cosmic abundance of chemical elements were collected by the Norwegian geochemist Victor Goldschmidt. The general idea in this line of reasoning is that the nuclear species, or at least some of them, are the products of nuclear processes in the early phase of the universe. This was the guiding philosophy of Gamow and his associates Ralph Alpher and Robert Herman, who in the late 1940s developed it into a research programme sometimes known as 'nuclear archaeology' (Kragh 1996, pp. 122-132). The apt phrase underlines the methodological similarity between this area of physical cosmology and ordinary historical archaeology. It refers to attempts to reconstruct the history of the universe by means of hypothetical cosmic or stellar processes and to test these by the resulting pattern of element abundances. Gamow was unable to account in this way for the heavier elements, but in collaboration with Alpher and Herman he succeeded in calculating the amount of helium in the universe to about 30% by weight, in reasonable agreement with observations. This early success of the big bang hypothesis was later much improved and extended to other light isotopes such as deuterium.

What matters is that by the late 1960s there was solid empirical evidence for the hot big bang, primarily in the form of the microwave background and the abundance of helium. This does not amount to a 'proof' of the big bang, but it does provide convincing evidence that makes it rational to accept the big bang picture (which does not imply that it is irrational not to accept it). Alternative cosmological models must, as a minimum, reproduce the empirical successes of the standard big bang model and do it without assumptions of an ad hoc nature. To do so on the basis of non-big bang assumptions turns out to be exceedingly difficult. It was the main reason why the steady state model of the universe was abandoned in the late 1960s. The lack of successful rival models is yet another reason to have confidence in the big bang, if by no means to accept it as true in a literal sense.

Whether students follow an evidence-based approach that corresponds to the historical development or not, it is not enough that they can justify their belief in the big bang picture in terms of evidence for it. They also need to know what this picture is, more exactly. If not the students will believe in the big bang, knowing why they believe it but not knowing what they believe in. Several studies show that students have quite different views of the nature of the origin and evolution of the universe. According to a study of Swedish upper-secondary students of age 18-19 years, they conceive the big bang in a variety of ways:

> For example, there are students saying that the universe has always existed in some way. Others talk about a beginning with the Big Bang, but show that they do not view this as an absolute beginning of the universe. … In addition to the view ascribed above where the Big Bang is viewed as something happening to the whole of the universe, there are also some students who talk about the Big Bang as the origin of the earth and/or the sun. (Hansson & Redfors 2006, p. 359)

One of the students described the big bang as an event 'where an explosion made gases and particles spread out in space and then they attracted each other and formed suns' (p. 366). Studies show consistently that the most common misconception of the big bang is to associate it with an explosion of pre-existing matter into empty space (Prather et al. 2003, Wallace et al. 2012, Bailey et al. 2012). Perhaps more surprisingly, only relatively few students connect the big bang to the beginning of the cosmic expansion, and very few think of it as an explosion from nothing.

Although it is hard not to think of the big bang as some kind of explosion of pre-existing matter, it is important to make the students understand that this is at best a somewhat flawed metaphor. Incidentally, Lemaître himself used the metaphor as early as 1931, when he spoke of his new big bang model as a 'fireworks theory', thereby trying to visualize what happened in the cosmic past. Fireworks explode into the surrounding air, but there is nothing 'outside' that the universe can explode into. While an explosion occurs at some location, the bang of the past did not happen somewhere in the universe. It was the entire universe that 'exploded' and thus the big bang happened everywhere. If this is hard to visualize, it is because it cannot be visualized.

It is also important to be aware that the qualitative meaning of the big bang is that long ago all distances, as given by the scale factor $R(t)$, were nearly zero, after which $R(t)$ increased rapidly. For some 14 billion years ago the universe was very compact, very hot and, in a sense, very small. The essence of the big bang is not a claim of an absolute beginning in some 'singularity' at $t = 0$, but a claim of a state of the universe, much earlier than and very different from the present state, that over long spans of time has evolved into the one we now observe. Another way of putting it is that the presently observed expansion started at some finite time ago in the cosmic past, so that the expanding universe can be ascribed a finite age. Note that this does not necessarily imply that the universe has a finite age (see also Sect. 6.4). Creation in an absolute and therefore metaphysical sense is not – and fortunately not – a part of the big bang scenario, just as little as an absolute origin of life is a necessary part of the neo-Darwinian evolution scenario.

**5. The Concept of the Universe**

Although cosmology has undoubtedly developed into a proper and impressive physical science since the 1960s, it is not just another branch of physics or astronomy. Nor is it just astrophysics extended from the stars to the universe at large. No, it is a very special and potentially problematic science in which questions of a philosophical (and sometimes religious) nature cannot be clearly separated from scientific questions relating to observation and theory. To present cosmology to students without taking into regard its special nature is to present them with a narrow and distorted picture of the fascinating science of the universe. Questions of a philosophical nature are part and parcel of what cosmology is about, and they should be given due consideration also in educational contexts, if not at the expense of the scientific issues. This is a major reason why modern cosmology, including aspects of its history, should have a prominent role in science teaching and why it enters significantly in many courses for students not majoring in physics or astronomy.

*5.1. The Cosmological Principle*

Much of cosmology's special and potentially problematic nature is independent not only of the big bang but also of the expansion of the universe. Indeed, being basically of a conceptual nature it is largely independent of modern scientific discoveries. A key problem, no less important today than it was in the time of Aristotle, is simply the unique domain of cosmology, this most peculiar concept of *the universe*. The standard definition of cosmology is something like 'the science of the universe', yet it is far from obvious that such a frightening concept as the universe can be the subject of scientific study. The relatively recent recognition that this can be done, and that even the universe at large is not foreign land to science, is one of the marvels of the modern physical sciences.

Among the epistemic problems that face a science of the universe is that cosmological knowledge seems to be conditioned by certain principles or assumptions that are completely unverifiable and for this reason may appear to be metaphysical rather than physical (Ellis 1984). The best known of these principles is the so-called cosmological principle, namely the generally held assumption that the universe is homogeneous and isotropic on a very large scale. It is sometimes referred to as the extended Copernican principle, a rather unfortunate name given that Copernicus' universe had the Sun as its fixed and unique centre. First explicitly formulated in 1932, the cosmological principle lies at the heart of all relativistic standard models, but it is not restricted to models governed by the general theory of relativity. Indeed, when British cosmologist Edward Milne introduced it in 1932, it was in connection with his own theory of the expanding universe which was entirely different from the theory governed by general relativity. The principle assumes that the vast ocean of unobservable regions of the universe is similar to the region we have empirical access to, a region that may well be an infinitesimal part of the entire universe. What is the epistemic status of the cosmological principle? Is it a necessary precondition for cosmology, or is it merely a convenience that may be accepted or not?

    The cosmological principle does have an empirical basis in so far that it roughly agrees with observations, but observations can say nothing about the structure of the universe far beyond the Hubble region, not to mention the cosmic horizon. Extrapolations much beyond this scale are necessarily hypothetical as they rest on an assumption of global uniformity that can never be verified. One might also say that they rest on 'faith', although the faith in the global validity of the cosmological principle is supported by local observations and therefore quite different from 'blind faith'. If cosmology rests on an unverifiable and perhaps metaphysical principle, can it still claim to be scientific? This is not to suggest that

the cosmological principle is in fact metaphysical, but to suggest that it is worth contemplating the status of the principle and to discuss it also in a teaching context rather than merely present it as a reasonable if unprovable assumption (Kuhn 1998, p. 551).

The instinct of many students majoring in science is to react with hostility and distrust to terms such as 'faith' and 'metaphysics'. (For students not majoring in science, see Shipman et al. 2002). Yet, because something is ultimately a matter of faith it does not imply that it is irrational, unscientific or arbitrary. There is an element of belief in most scientific ideas. It is important to recognize that unverifiability is not a great methodological sin that automatically deprives a theory or field its scientific status. In fact, students are well aware of high-status scientific theories that cannot be verified, although they may never have thought of them as theories that, in a manner of speaking, rest on belief.

Several of our commonly accepted laws of physics can be said to be cosmological in nature in so far that they are claimed to be true all over the universe and in any patch of cosmic space-time. Newton's law of gravitation speaks of the attractive force between any two masses in the universe, and the law of energy conservation is valid for all processes at any time in the universe. They can reasonably be considered statements relating to the universe at large and for this reason implicitly of a cosmological nature. Of course, neither these two laws nor other similar laws can be verified experimentally. The moral is that students have no reason to fear unverifiability in cosmology, since we have to live with this feature anyway. On the other hand, unfalsifiability is a different matter.

Contrary to what some philosophers have argued (Munitz 1986), the cosmological uniformity principle and similar principles are not of an a priori nature, that is, true by necessity. The cosmological principle is a simplifying assumption that could be proved wrong by observation. In that case it would have to be abandoned, but this would not make cosmology impossible, only more complicated. There are plenty of theoretical cosmological models that do not presuppose homogeneity or isotropy. The case exemplifies the important distinction between verifiability and falsifiability that is a central message in Popperian philosophy of science. That global uniformity principles of this kind are indeed falsifiable is further illustrated by the 'perfect cosmological principle' upon which the now defunct steady state theory was based. This principle extended the cosmological principle to the temporal dimension, namely, by claiming that there is no privileged time in the history of the universe any more than there is a privileged position. When the steady state theory was put in the grave in the 1960s, so was the perfect cosmological principle.

*5.2. The Uniqueness of the Universe*

The universe does not only stretch beyond the observable region, it is also, at least according to the ordinary meaning of the term, a unique concept (Ellis 1999). If the universe by definition comprises everything of a physical nature, space and time included, there can only be one universe. Contrary to ordinary physics, which operates with objects and phenomena which are local and of which there are many, the universe is not a member or instance of a class of objects. Newton could establish his inverse-square law of gravitation because there are many bodies that gravitate. By observing and experimenting with different initial conditions he and later physicists could confirm the validity of the law. But not so with respect to the universe, where the initial conditions are fixed and unchangeable. We cannot re-run the universe with the same or altered conditions to see what would happen if they were different. It seems to follow that we cannot establish proper cosmological laws *of* the universe comparable to the ordinary laws of physics, for we cannot test any such proposed law except in terms of being consistent with a singular 'object', the observed universe.

Since we use laws to explain things, such as explaining the falling apple as an instance of the law of gravity or the energy generated by the Sun as an instance of the laws of quantum physics, it may seem that the domain of cosmology is beyond explanation in the causal-nomological sense normally used in physics. To put it differently, whereas in local physics law-governed and contingent properties can be distinguished, this may not be possible in cosmology. Does it follow that the universe – the domain of cosmology – is beyond explanation? The question was discussed by René Descartes and his contemporaries in the seventeenth century, and it has continued to attract attention from both philosophers and cosmologists. According to Descartes, the divine mechanical laws guaranteed that the original chaos, whatever its structure and initial conditions, would evolve into our universe or one indistinguishable from it. Newton, on the other hand, insisted that the universe cannot be fully understood by the laws of mechanics alone. Descartes' 'indifference principle' continues to play a role in modern cosmology, except that the laws are no longer seen as mechanical only (McMullin 1993).

There are ways to avoid the pessimistic conclusion that the universe is beyond explanation. One strategy is simply to deny the uniqueness of the universe by postulating the existence of many others (Sect. 6.1). Another solution is to recall that there are other forms of explanation than those used in the standard deductive-nomological scheme. Because cosmology is a non-nomological science, it does not follow that it is impossible to account for the present state of the universe. Thus, to explain the fact that the present temperature of the microwave background is about 2.7 K we do not need a law of the universe or an ensemble of universes we can compare ours with. We can and do offer an explanation – not a causal one, but a historical or genetic explanation – by accounting for how the background radiation cooled with the expansion of the universe.

**6. Unfinished Businesses**

Cosmology of the twenty-first century is in some respects an unfinished business that may provide students with a rare insight in science in vivo. Not only are there important scientific questions that are not solved yet, most notably the nature of dark matter and dark energy, there are also questions of old vintage that may belong as much to philosophy as to science and about which we do not even know whether they are answerable or not. Many students are naturally curious about the kind of borderline questions that cosmology present us with, and teachers should do what they can to satisfy their curiosity. Students should be confronted with problems of this kind and be stimulated to think about them in a critical and rational way. They should not be dissuaded from asking questions even though these may appear to be naïve – maybe they are not so naïve after all. Modern physical cosmology is a wonderful resource for enlightenment and discussion of questions that relate to the limits of science. Contrary to what is the case in most other sciences, such questions are integrated parts of the science of the universe understood broadly. In general science courses dealing with cosmology it will be natural to introduce at least some of the issues.

*6.1. Many Universes?*

A typical textbook definition is that 'the visible universe is the largest astronomical structure of which we have any knowledge' (Arny 2004, p. 9). This is a reasonable and operational definition, but why restrict cosmology to the study of the visible universe? There surely is something behind it. In the more general and ambitious sense adopted by some cosmologists the universe is taken to be 'everything that exists'. If so, it makes no sense to speak of other universes. Nonetheless, this is what several theoretical cosmologists do nowadays, where the question of the definition of the universe has been reconsidered as part of the controversy over the 'multiverse', the hypothesis that there is a multitude of different universes of which the one we observe is only a single member (Carr & Ellis 2008, Kragh 2011b). This ongoing controversy has many interesting aspects, not least that critics have questioned the scientific nature of the multiverse hypothesis and thus reopened the old question of whether cosmology, or some versions of cosmology, belongs to physics or metaphysics. On the other hand, advocates of the multiverse argue that it is a scientific idea and that it follows from, or is strongly suggested by, recent developments within string theory and inflation cosmology. Although the multiverse cannot be tested directly, they claim that it leads to testable consequences.

The existence of a cosmic horizon beyond which we will never be able to see or otherwise get information from, not even in principle, is not a new insight. As early as 1931 Eddington pointed out that the accelerated expansion of the closed Lemaître-Eddington universe would eventually lead to 'a number of disconnected universes no longer bearing any physical relation to one another' (Eddington 1931, p. 415). This kind of multiverse is relatively innocent, since the different universes, although causally separated, inhabit the same space-time. More extreme and more speculative is the modern idea of a huge number of disparate universes, each of them with its own physical laws and constants of nature (and with ours being perhaps the only one with intelligent life). We obviously cannot have empirically based knowledge about the content and properties of these other worlds, nor can we establish their existence observationally. The numerous other worlds may exist or not, but if the question cannot be decided by means of experiment and observation does it belong to science?

The recent controversy over the universe may well be used in the teaching of introductory cosmology as it does not rely on advanced theories but is essentially of a qualitative and philosophical nature. A recommendable source, most relevant also for the purpose of teaching, can be found in a discussion between George Ellis and Bernard Carr in the journal *Astronomy & Geophysics* (Carr & Ellis 2008). This illuminating source has for some years been used in courses in philosophy of science for undergraduate science students at Aarhus University, and with considerable success. It works very well and provokes much good discussion among the students.

*6.2. Infinite Space*

The problem of the spatial extension of the universe is another of those cosmological questions that have been discussed since Greek antiquity and that we still do not know the answer to. While Einstein's original universe of 1917 was positively curved and with a definite volume, corresponding to a curvature radius of only about $10^7$ light-years, the expanding Einstein-de Sitter model of 1932 assumed a flat and therefore infinite space. The same was the case of the steady state universe, where a zero curvature parameter $k = 0$ follows from the perfect cosmological principle. Indirect and model-dependent measurements of the curvature of cosmic space did not lead to a definite answer, but the present consensus model (including inflation and dark energy) strongly favours a flat universe of infinite extent. Assuming the cosmological principle, this implies a universe with an infinite number of objects in it, whether these being electrons or galactic clusters.

Students may tend to think of infinity as just a hugely large number, but (as Newton was well aware of) there is a world of difference between the extremely large and the infinitely large. Actual infinities are notoriously problematical, leading to all kinds of highly bizarre consequences. The general attitude of modern cosmologists is to ignore the troublesome philosophical problems of actual infinities and speak of the infinite universe as just an indefinitely large universe, not unlike the students' intuition. Only rarely do they reflect on the weird consequences of the actual infinite – but perhaps they should. Ellis is one of the relatively few cosmologists who take the infinite cosmos seriously, suggesting that the infinities may not be real after all, indeed cannot be real. Ellis and his collaborators argue that physical quantities cannot be truly infinite and that infinite sets of astronomical objects have no place in cosmology. If such quantities formally turn up in a theory or model, it almost certainly means that the theory is wrong. Infinity, they emphasize, 'is not the sort of property that can be physically realized in an entity, an object, or a system, like a definite number can' (Stoeger, Ellis & Kirchner 2008, p. 17).

Although an infinite universe follows from some cosmological models, we will never know whether the universe is in fact infinite. Observations and theory indicate a flat space, but observations are limited to the visible universe. It is only by assuming the cosmological uniformity principle that we can extrapolate to the universe at large. Moreover, we can never know observationally whether $k = 0$ precisely, only that $k$ varies between the limits $\pm \Delta k$ corresponding to the inevitable observational uncertainties. This observational asymmetry between flat and curved space was pointed out by the Russian mathematician Nikolai Lobachevsky as early as 1829, a century before the expanding universe. It is worth noticing that although the idea of curved space only was adopted by physicists and astronomers with Einstein's general theory of relativity, as a mathematical idea it goes back to the first half of the nineteenth century.

*6.3. The Enigma of Creation*

The traditional version of the big bang theory inevitably invites questions of a philosophical and to some extent religious nature concerning the origin of everything. Although the big bang model is not really a model of absolute beginning or creation, but a cosmic evolutionary scenario, it would be artificial to ignore these questions and simply dismiss them as unscientific. Unscientific they may be, but they are no less natural and fascinating for that. Teachers can keep them out of astronomy and general science courses, but that would be to betray the curiosity and natural instincts of the students. Moreover, questions concerning cosmic creation have a long a glorious history which makes interesting connections between the history of science and the history of ideas, philosophy and religious thought. Whether one likes it or not, the creation of the physical universe is part of the world view of most cultures, and for this reason alone it should not be ignored in science courses. Fortunately there is a rich literature on philosophical, political and religious world views and their place in science education (Matthews 2009).

The problem with creation in a cosmological context is that if we conceive the big bang as an absolute beginning at $t = 0$, then a causal scientific explanation of the creation event is impossible. After all, a cause must come before the effect, and there is no 'before'. Current cosmology has traced the history of the universe back in time to the inflationary period which is supposed to have occurred at $t = 10^{-34}$ s or thereabout. It is often assumed that the cosmic past can be traced even farther back to the Planck time at $t = 10^{-43}$ s (and there are even speculative pre-Planck theories). But however close calculations may bring us to the magical moment $t = 0$, it seems in principle impossible to account for the creation event itself. To say that the universe was created in a space-time singularity is a mere play with words, since the singularity is mathematical abstraction devoid of physical content. Physics did not exist at $t = 0$ and it makes no sense to speak of physical mechanisms where even the concepts of cause and effect cannot be defined.

In spite of the rhetoric of some cosmologists, there are no scientific theories that explain the origin of the universe from 'nothing' and there never will be such theories. The concept of nothingness or absolute void has a rich history (Genz 1999) that recently has become relevant to science, not least after the discovery of the dark energy that is generally identified with the vacuum energy density as given by the cosmological constant and interpreted in terms of quantum mechanics. However, the modern quantum vacuum is entirely different from absolute nothingness. There cannot possibly be a scientific answer to what nothingness is, and yet it does not therefore follow that the concept is meaningless.

A major reason why big bang cosmology has been and to some extent still is controversial in the eyes of the public, is that it may be seen as a scientific version of Genesis or at least to provide scientific justification for a divinely created world. This misguided view was endorsed by pope Pius XII in 1951 (Kragh 1996, pp. 256-259) and is still popular in some circles. Although this is not the place to discuss the complex relations between cosmology and religion (Halvorsen & Kragh 2010), it appears that some of these questions are suited for discussions with and among students and should not necessarily be kept out of the physics classroom. Courses that aim to establish a dialogue between science and religion have existed for some time, and in some of them cosmology enters prominently (Shipman et al. 2002). The issue is also mentioned in Kuhn (1998), a textbook which includes a brief and admirably clear exposition of the relationship between cosmology and religious faith:

> If we use God as an explanation for the big bang, there would be no reason to look further for a natural explanation. Use of supernatural explanations would shut down science. … If science relied on a creator to explain the inexplicable, there would be nowhere to go, no way to prove that explanation wrong. The question

> would have already been settled. … Science does not deny the existence of God. God is simply outside its realm. (Kuhn 1998, p. 557)

While much attention is paid to the origin of the universe, the other end of the cosmic time scale is rarely considered a question of great importance. And yet Einstein's equations of relativistic cosmology are symmetric in time, telling us not only about the distant past but also about the remote future. Will the universe ever come to an end? If so, what kind of end? In the late nineteenth century these questions were eagerly discussed in relation to the so-called heat death supposedly caused by the increase of entropy in the universe, and recently they have been reconsidered within the framework of modern physics and cosmology. The new subfield known as 'physical eschatology' is concerned, among other things, with the final state of life and everything else (Kragh 2011b, pp. 325-353). Parts of physical eschatology are controversial and highly speculative, yet it is a subject that is likely to appeal to many students and that they should know about. As the birth of the universe relates to religious dogmas, so does it death.

*6.4. A Universe without a Beginning*

In his last book, *The Demon-Haunted World*, the prominent astronomer and science popularizer and educator Carl Sagan pointed out that science might conceivably demonstrate the universe to be infinitely old. He suggested that 'this is the one conceivable finding of science that could disprove a Creator – because an infinitely old universe would never have been created' (Sagan 1997, p. 265). On the face of it Sagan's assertion may appear convincing, perhaps even self-evident, but it is based on a misunderstanding that conflates the scientific notion of 'finite age' with the theological notion of 'creation'. Theologians and Christian philosophers agree that even an infinitely old universe would have to be created, in the sense of being continuously sustained, and that it would in no way pose problems for faith. Even if the universe had existed in an infinity of time, we could still ask for the reason of its existence, or why it was created.

We have very good reason to believe in the big bang, but we have no good reason to believe that this is how the universe ultimately came into being. Concepts such as cosmic origin and time are difficult, not only conceptually but also for semantic reasons. Thus, we would presumably think that whereas the steady state universe of Hoyle and others had always existed, this is not the case with the finite-age big bang universe. The two statements 'the universe has a finite age' and 'the universe has always existed' appear to be contradictory, but in reality they may both be true. To say that the universe has always existed is to say that it existed whenever time existed. The word 'always' is a temporal term that presupposes time. Since it is hard to imagine time without a universe, it makes sense to speak of a big bang universe which has always existed. The phrase 'the universe has always existed' reduces to a tautology. This observation is more than just a philosophical nicety, as illustrated by one of the questions posed to students in a questionnaire: 'Does the universe have an age, or has it always existed' (Bailey et al. 2012). Several of the students, we are told, 'gave a contradictory response, such as "the universe has always existed: it is billions of years old"'. As argued, the answer is not really contradictory.

Until recently it was taken for granted that a universe of finite age implies an absolute cosmic beginning of some kind. The traditional answer to the supposedly naïve question of what there was before the beginning in the big bang has been to dismiss or ridicule it as an illegitimate and meaningless question. For how can there be something 'before' the beginning of time? But there is no reason to ridicule the question if it is recognized that the big bang event at $t = 0$ need not mark the beginning of time.

During the last two decades an increasing number of cosmologists have argued that the big bang picture does not preclude a past eternity in the form of, for example, one or more earlier universes. Most theories of quantum gravity operate with a non-singular smallest volume, which makes it possible to extend cosmic time through the $t = 0$ barrier at least in a formal sense.[5] There exists presently a handful of such theories, which are all speculative to varying degrees but nonetheless are considered serious scientific hypotheses. To mention but one example, according to so-called loop quantum cosmology the universe was not created a finite time ago but exists eternally. There was a big bang, of course, but in the form of a well-described transition of the universe from a contracting to an expanding phase. The space of loop quantum cosmology is discrete on a very small scale, which has the observable consequence that photons of very high energy should travel faster than those of low energy.

Did the universe have an absolute beginning in time or not? The most honest answer is probably that we do not know and perhaps cannot ever know. It may be one of those questions about which we cannot even tell whether it is meaningful or not or whether it belongs to science or not.

## 7. Conclusion

The cosmological world view of the twenty-first century, largely identical to the standard big bang theory, is to a considerable extent what the Copernican world system was in the seventeenth century. Just as this system was not only a new theory of astronomy, but also carried with it wider implications related to philosophy, religion and social order, so the modern picture of the universe cannot be easily separated from extra-scientific considerations. Such considerations, be they of a philosophical, conceptual or religious nature, should to some extent appear also in the teaching of science and do it in a qualified and critical manner.

---

[5] It is far from obvious that the symbol $t$, as it appears in the equations describing the very early universe near or before the Planck time $t = 10^{-43}$ s, can be ascribed a well-defined physical meaning (Rugh & Zinkernagel 2009). The meaning of time is even less clear in theories of quantum cosmology describing the universe before $t = 0$.

One of the important aims of science education is to bring home the lesson that although science provides us with reliable and privileged knowledge of nature, it does not answer all questions that are worth asking. This lesson emerges with particular force from the study of cosmology. It may be expressed more poetically with a famous quotation from Shakespeare's *Hamlet*: 'There are more things in heaven and Earth, Horatio, than are dreamt of in your philosophy'. Recall that at the time of Shakespeare, the term 'philosophy' had a meaning corresponding to our 'science'.